\pdfoutput = 1
\documentclass[12pt,a4paper]{article}

\usepackage{lineno}  
\usepackage{graphicx}  
\usepackage{xspace} 
\usepackage{color}
\usepackage{colortbl}
\usepackage{amsmath} 
\usepackage{ifthen} 

\newboolean{pdflatex}
\setboolean{pdflatex}{true} 

\newboolean{articletitles}
\setboolean{articletitles}{true} 

\newboolean{uprightparticles}
\setboolean{uprightparticles}{true} 

\usepackage{amssymb}
\usepackage{amsfonts}
\usepackage{upgreek} 

\usepackage{graphpap}
\usepackage{multirow}
\usepackage{rotating}

\usepackage{hyperref}    
\usepackage[all]{hypcap} 

\def\lhcb {LHCb\xspace}
\def\ux85 {UX85\xspace}

\def\spd    {SPD\xspace}
\def\presh  {PS\xspace}

\ifthenelse{\boolean{uprightparticles}}%
{
 
 \def\Pgamma      {\ensuremath{\upgamma}\xspace}

 \def\Peta        {\ensuremath{\upeta}\xspace}

 \def\Pmu         {\ensuremath{\upmu}\xspace}

 \def\Ppi         {\ensuremath{\uppi}\xspace}                 
                  
 \def\Prho        {\ensuremath{\uprho}\xspace}

 \def\Ppsi        {\ensuremath{\uppsi}\xspace}

 \def\PDelta      {\ensuremath{\Delta}\xspace}                 
 \def\PXi      {\ensuremath{\Xi}\xspace}                 
 \def\PLambda      {\ensuremath{\Lambda}\xspace}                 
 \def\PSigma      {\ensuremath{\Sigma}\xspace}                 
 \def\POmega      {\ensuremath{\Omega}\xspace}                 
 \def\PUpsilon      {\ensuremath{\Upsilon}\xspace}                 
 
 \def\PB      {\ensuremath{\mathrm{B}}\xspace}                 
                  
 \def\PD      {\ensuremath{\mathrm{D}}\xspace}

 \def\PJ      {\ensuremath{\mathrm{J}}\xspace}                 
 \def\PK      {\ensuremath{\mathrm{K}}\xspace}

 \def\Pb      {\ensuremath{\mathrm{b}}\xspace}                 
 \def\Pc      {\ensuremath{\mathrm{c}}\xspace}

 \def\Pi      {\ensuremath{\mathrm{i}}\xspace}

}
{
 
 \def\Pgamma      {\ensuremath{\gamma}\xspace}

 \def\Peta        {\ensuremath{\eta}\xspace}

 \def\Pmu         {\ensuremath{\mu}\xspace}

 \def\Ppi         {\ensuremath{\pi}\xspace}                 
                  
 \def\Prho        {\ensuremath{\rho}\xspace}

 \def\Ppsi        {\ensuremath{\psi}\xspace}                 
                  
 \mathchardef\PDelta="7101
 \mathchardef\PXi="7104
 \mathchardef\PLambda="7103
 \mathchardef\PSigma="7106
 \mathchardef\POmega="710A
 \mathchardef\PUpsilon="7107
                  
 \def\PB      {\ensuremath{B}\xspace}                 
                  
 \def\PD      {\ensuremath{D}\xspace}

 \def\PJ      {\ensuremath{J}\xspace}                 
 \def\PK      {\ensuremath{K}\xspace}

 \def\Pb      {\ensuremath{b}\xspace}                 
 \def\Pc      {\ensuremath{c}\xspace}

 \def\Pi      {\ensuremath{i}\xspace}

}

\def\mumu       {\ensuremath{\Pmu^+\Pmu^-}\xspace}

\def\g      {\ensuremath{\Pgamma}\xspace}

\def\cquark    {\ensuremath{\Pc}\xspace}

\def\bquark    {\ensuremath{\Pb}\xspace}

\def\pion  {\ensuremath{\Ppi}\xspace}
\def\piz   {\ensuremath{\pion^0}\xspace}

\def\pip   {\ensuremath{\pion^+}\xspace}

\def\kaon  {\ensuremath{\PK}\xspace}
  \def\Kbar  {\kern 0.2em\overline{\kern -0.2em \PK}{}\xspace}

\def\Kz    {\ensuremath{\kaon^0}\xspace}
\def\Kzb   {\ensuremath{\Kbar^0}\xspace}
\def\KzKzb {\ensuremath{\Kz \kern -0.16em \Kzb}\xspace}
\def\Kp    {\ensuremath{\kaon^+}\xspace}
\def\Km    {\ensuremath{\kaon^-}\xspace}

\def\KpKm  {\ensuremath{\Kp \kern -0.16em \Km}\xspace}

  \def\Dbar    {\kern 0.2em\overline{\kern -0.2em \PD}{}\xspace}
\def\D       {\ensuremath{\PD}\xspace}

\def\Dz      {\ensuremath{\D^0}\xspace}
\def\Dzb     {\ensuremath{\Dbar^0}\xspace}
\def\DzDzb   {\ensuremath{\Dz {\kern -0.16em \Dzb}}\xspace}
\def\Dp      {\ensuremath{\D^+}\xspace}
\def\Dm      {\ensuremath{\D^-}\xspace}

\def\DpDm    {\ensuremath{\Dp {\kern -0.16em \Dm}}\xspace}

\def\B       {\ensuremath{\PB}\xspace}
  \def\Bbar    {\kern 0.18em\overline{\kern -0.18em \PB}{}\xspace}

\def\Bu      {\ensuremath{\B^+}\xspace}

\def\jpsi     {\ensuremath{{\PJ\mskip -3mu/\mskip -2mu\Ppsi\mskip 2mu}}\xspace}

  \def\Y#1S{\ensuremath{\PUpsilon{(#1S)}}\xspace}

\def\photos     {\mbox{\textsc{Photos}}\xspace}

\def\BF         {{\ensuremath{\cal B}\xspace}}

\def\BR         {\BF}

\def\to                 {\ensuremath{\rightarrow}\xspace}

\def\AT#1     {\ensuremath{A_{\mathrm{T}}^{#1}}\xspace}           

\def\C#1      {\ensuremath{\mathcal{C}_{#1}}\xspace}                       
\def\Cp#1     {\ensuremath{\mathcal{C}_{#1}^{'}}\xspace}                    
\def\Ceff#1   {\ensuremath{\mathcal{C}_{#1}^{\mathrm{(eff)}}}\xspace}        
\def\Cpeff#1  {\ensuremath{\mathcal{C}_{#1}^{'\mathrm{(eff)}}}\xspace}       
\def\Ope#1    {\ensuremath{\mathcal{O}_{#1}}\xspace}                       
\def\Opep#1   {\ensuremath{\mathcal{O}_{#1}^{'}}\xspace}                    

\newcommand{\tev}{\ensuremath{\mathrm{\,Te\kern -0.1em V}}\xspace}
\newcommand{\gev}{\ensuremath{\mathrm{\,Ge\kern -0.1em V}}\xspace}
\newcommand{\mev}{\ensuremath{\mathrm{\,Me\kern -0.1em V}}\xspace}
\newcommand{\kev}{\ensuremath{\mathrm{\,ke\kern -0.1em V}}\xspace}
\newcommand{\ev}{\ensuremath{\mathrm{\,e\kern -0.1em V}}\xspace}
\newcommand{\gevc}{\ensuremath{{\mathrm{\,Ge\kern -0.1em V\!/}c}}\xspace}
\newcommand{\mevc}{\ensuremath{{\mathrm{\,Me\kern -0.1em V\!/}c}}\xspace}
\newcommand{\gevcc}{\ensuremath{{\mathrm{\,Ge\kern -0.1em V\!/}c^2}}\xspace}
\newcommand{\gevgevcccc}{\ensuremath{{\mathrm{\,Ge\kern -0.1em V^2\!/}c^4}}\xspace}
\newcommand{\mevcc}{\ensuremath{{\mathrm{\,Me\kern -0.1em V\!/}c^2}}\xspace}

\def\mum  {\ensuremath{\,\upmu\rm m}\xspace}

\newcommand{\stat}{\ensuremath{\mathrm{\,(stat)}}\xspace}
\newcommand{\syst}{\ensuremath{\mathrm{\,(syst)}}\xspace}

\newcommand{\chisq}{\ensuremath{\chi^2}\xspace}

\def\gsim{{~\raise.15em\hbox{$>$}\kern-.85em
          \lower.35em\hbox{$\sim$}~}\xspace}
\def\lsim{{~\raise.15em\hbox{$<$}\kern-.85em
          \lower.35em\hbox{$\sim$}~}\xspace}

\def\ptot       {\mbox{$p$}\xspace}

\def\evtgen     {\mbox{\textsc{EvtGen}}\xspace}
\def\pythia     {\mbox{\textsc{Pythia}}\xspace}

\def\geant      {\mbox{\textsc{Geant4}}\xspace}

\def\tell1  {TELL1\xspace}
\def\ukl1   {UKL1\xspace}

\usepackage{mciteplus}
\usepackage[noadjust]{cite}
\usepackage{filecontents}

\begin{document}

\pagestyle{plain} 
\setcounter{page}{1}
\pagenumbering{arabic}


\vspace*{-1.5cm}
\hspace*{-0.5cm}
\noindent
\begin{tabular*}{\linewidth}{lc@{\extracolsep{\fill}}r@{\extracolsep{0pt}}}
\ifthenelse{\boolean{pdflatex}}
{\vspace*{-2.7cm}\mbox{\!\!\!\includegraphics[width=.14\textwidth]{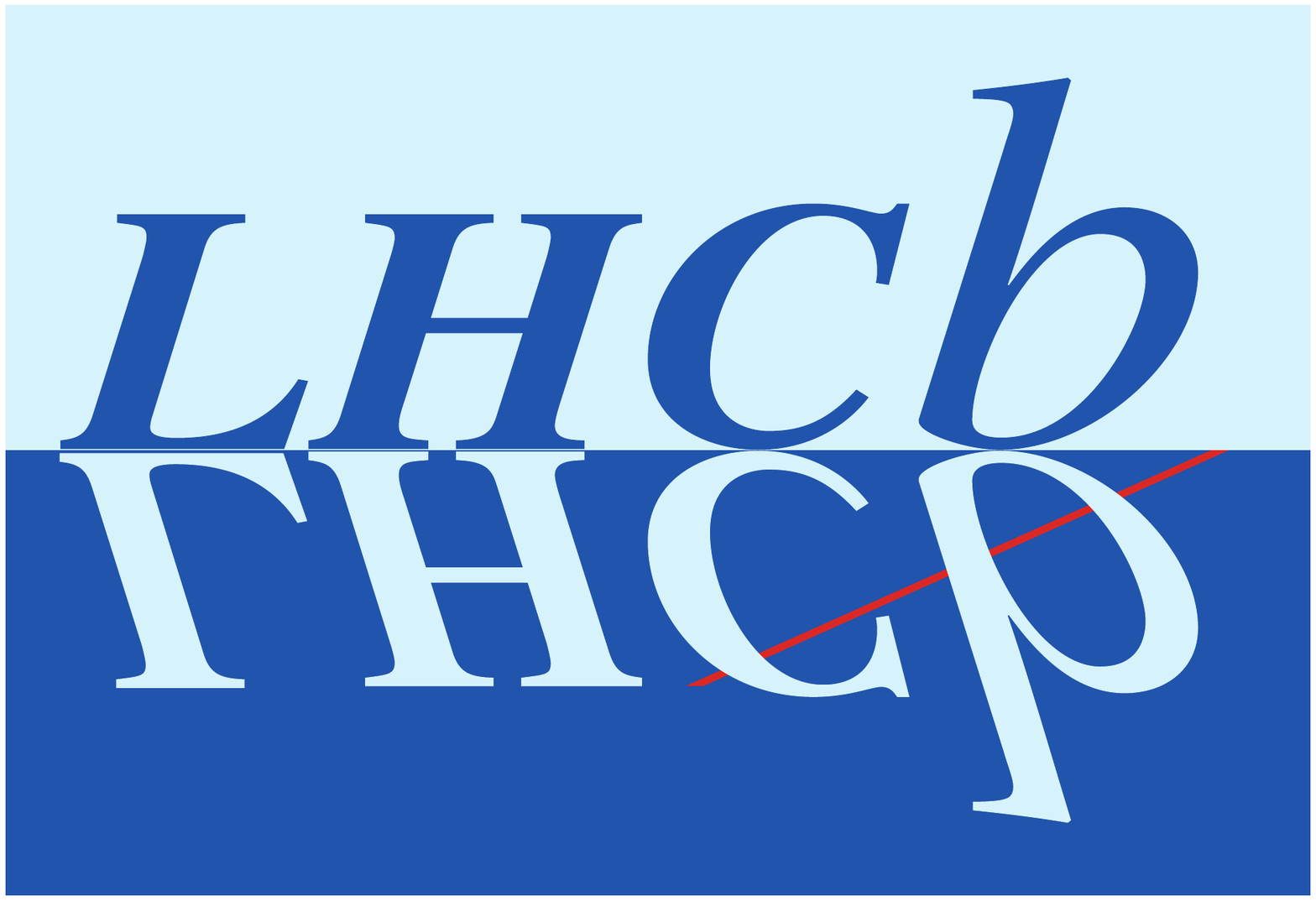}} & &}%
{\vspace*{-1.2cm}\mbox{\!\!\!\includegraphics[width=.12\textwidth]{lhcb-logo.eps}} & &} \\
 & & CERN-LHCb-PROC-2015-009 \\  
 & & \today \\ 
 & & \\
\hline
\end{tabular*}

{\bf\boldmath\huge
\begin{center}
Study of photon reconstruction~efficiency using \mbox{$\Bu\to\jpsi{}\mathrm{K}^{(*)+}$}~decays  
in the LHCb experiment
\end{center}
}

\begin{center}
Ekaterina Govorkova$^{1,2}$
\bigskip\\
{\it\footnotesize
$^1$Institute for Theoretical and Experimental Physics, ITEP, Moscow, Russia\\
$^2$Skobeltsyn Institute of Nuclear Physics, Moscow State University, Moscow, Russia \\ \smallskip 
}
{\footnotesize The proceedings of the 18th International Moscow School of Physics (43d ITEP Winter School) \\ }
\end{center}

\begin{abstract}
  \noindent
  The reconstruction efficiency of photons and neutral pions is measured using the relative yields of reconstructed \mbox{$\Bu\to\jpsi{}\mathrm{K}^{*+}\left(\to\mathrm{K}^+\piz\right)$}~and \mbox{$\Bu\to\jpsi{}\mathrm{K}^+$}~decays. The efficiency is studied using the data set, corresponding to an integrated luminosity of $3~\mathrm{fb}^{-1}$, collected by the LHCb experiment in proton-proton collisions at the centre-of-mass energies of~\mbox{7 and 8 TeV}.   
\end{abstract}

\section{Introduction}
\label{sec:intro}

The \lhcb detector~\cite{Alves:2008zz} is a single-arm forward spectrometer at LHC covering the \mbox{pseudorapidity} range $2<\eta <5$, designed for the study of particles containing \bquark or \cquark quarks. Many of the decays under study have photons and neutral pions in the final state, for example~\cite{Aaij:1692766,Aaij:1484958,Aaij:1966993}. The knowledge of photon reconstruction efficiency in a wide range of photon kinematic is a crucial ingredient for such kind of measurement~\cite{Adeva:2009ny},~\cite{Aaij:2014hla}. This manuscript describes the study of photon reconstruction efficiency using large sample of \mbox{$\Bu\to\jpsi{}\mathrm{K}^{*+}$}~decays and known ratio of branching fractions of \mbox{$\Bu\to\jpsi{}\mathrm{K}^{*+}$}~and \mbox{$\Bu\to\jpsi{}\mathrm{K}^+$}~decays~\cite{PDG-2014}. A correction factor for the reconstruction efficiency gained from simulation, is obtained through the detailed comparison between simulation and data.

The photon reconstruction efficiency is determined in data using comparison between \mbox{$\Bu\to\jpsi{}\mathrm{K}^{*+}\left(\to\mathrm{K}^+\piz\right)$} and \mbox{$\Bu\to\jpsi\mathrm{K}^{+}$} signal yields. The correction factor to \piz~reconstruction efficiency is obtained as the following:
\begin{equation}
\Peta^{\mathrm{corr}}_{\piz} = 
\dfrac{  N^{\Bu\to\jpsi{}\mathrm{K}^{*+}\left(\to\mathrm{K}^+\piz\right)}}
      {  N^{\Bu\to\jpsi{}\mathrm{K}^{+}}} 
\times  
\dfrac {  \varepsilon^{\mathrm{MC}}_{\Bu\to\jpsi{}\mathrm{K}^{+}}} 
       {  \varepsilon^{\mathrm{MC}}_{\Bu\to\jpsi{}\mathrm{K}^{*+}\left(\to\mathrm{K}^+\piz\right)}}
\times  
\dfrac {  \BR\left( \Bu\to\jpsi{}\mathrm{K}^{+}\right)} 
       {  \BR\left( \Bu\to\jpsi{}\mathrm{K}^{*+}\left(\to\mathrm{K}^+\piz\right)\right) },
\label{eq:main}
\end{equation}
where $N$~is the event yield, $\varepsilon^{\mathrm{MC}}$~is the total efficiency, determined using simulation, and the ratio of branching fractions is known to be~\cite{PDG-2014}
\begin{equation*}
\dfrac {  \BR\left( \Bu\to\jpsi{}\mathrm{K}^{+}\right)}
       {  \BR\left( \Bu\to\jpsi{}\mathrm{K}^{*+}\left(\to\mathrm{K}^+\piz\right)\right) }
 = 
\left( \frac{1}{3} \times \left( 1.39 \pm 0.09 \right) \right)^{-1}. 
\label{eq:ratio}
\end{equation*}
Since these decays have similar topology, the large cancellation of various systematic uncertanties, after imposing similar selection and trigger criteria is expected.

\section{LHCb detector and data sets}
\label{sec:detect}
The~\lhcb~detector includes a high-precision tracking system consisting of a silicon-strip vertex detector surrounding the pp interaction region~\cite{LHCbVELOGroup:2014uea}, a large-area silicon-strip detector located upstream of a~dipole magnet with a bending power of about $4{\rm\,Tm}$, and three stations of silicon-strip detectors and straw drift tubes~\cite{Arink:2013twa} placed downstream of the~magnet. The~tracking system provides a measurement of momentum, \ptot,  with a relative uncertainty that varies from 0.5\% at low momentum to 1.0\% at 200\gevc. Different types of charged hadrons are distinguished using information from two ring-imaging Cherenkov detectors~\cite{Adinolfi:1495721}. Photon, electron and hadron candidates are identified by a calorimeter system consisting of a scintillating-pad detector\,(SPD) which is designed to identify charged particles, followed by a wall of lead and the preshower detectors\,(PS), an electromagnetic calorimeter~(ECAL) and a hadronic calorimeter~(HCAL). Muons are identified by a system composed of alternating layers of iron and multiwire proportional chambers~\cite{LHCb-DP-2012-002}. 

The calorimeter system performs several tasks, providing the first level trigger with high transverse momentum photon, electron and hadron candidates, measuring their energies and positions and performing the separation between photons, electrons and hadrons. All calorimeters use the same energy detection principle: scintillation light is transmitted to a photomultiplier tube (PMT) by wavelength-shifting (WLS) fibres. The single fibres for the SPD/PS cells are read out using multianode photomultiplier tubes (MAPMT), while the fibre bunches in the ECAL and HCAL modules require individual phototubes. The SPD/PS, ECAL and HCAL adopt different cell dimensions in the inner and outer parts of the calorimeter since the hit density varies by two orders of magnitude over the calorimeter surface. A segmentation into three different sections has been chosen for the ECAL and projectively for the SPD/PS. Given the dimensions of the hadronic showers, the HCAL is segmented into two zones with larger cell sizes. 

The electromagnetic calorimeter employs ''shashlik'' technology of alternating scintillating tiles and lead plates. The energy resolution of the ECAL modules was determined at the test beam. The parametrisation $\sigma_{\mathrm{E}}/\mathrm{E} = \mathrm{a}/\sqrt{\mathrm{E}}\oplus \mathrm{b}\oplus \mathrm{c}/\mathrm{E}$ (E in GeV) is used, where a, b and c stand for the stochastic, constant and noise terms respectively. Depending on the type of module and test beam conditions the stochastic and constant terms were measured to be $8.5\% < \mathrm{a} < 9.5\%$ and b $\approx0.8\%$. The noise term corresponds to c = 0.003 GeV.

Energy deposits in ECAL cells are clusterised applying a $3\times3$ cell pattern around local maxima of energy deposition. Consequently the centers of the reconstructed clusters are always separated at least by one cell. If one cell is shared between several reconstructed clusters, the energy of the cell is redistributed between the clusters proportionally to the total cluster energy. The process is iterative and it converges rapidly due to relatively small ratio of the Moli\`ere radius (3.5 cm) to the cell size. After the redistribution of energy of shared cells, the energy-weighted cluster moments up to the order 2 are evaluated to provide the (hypotheses-independent) cluster parameters, namely, the total energy, the energy barycenter position and the two-dimensional energy spread matrix. Neutral clusters are identified as those clusters that do not match to charged tracks. The photon energy, $\mathrm{E}_{\mathrm{c}}$, is evaluated from the total cluster energy, $\varepsilon_{\mathrm{ECAL}}$, and the reconstructed energy deposit in the PS, $\varepsilon_{\mathrm{PS}}$, as follows: $\mathrm{E}_{\mathrm{c}} = \alpha \times \varepsilon_{\mathrm{ECAL}} + \beta \times \varepsilon_{\mathrm{PS}} $. The parameters $\alpha$ and $\beta$ account for energy leakage. The value of parameter $\alpha$ depends on the position of the cluster on the calorimeter surface while $\beta$ is estimated afterwards to give full account of energy samples~\cite{LHCb-2003-091, LHCb-2003-092}.

The calibration of the electromagnetic calorimeter is performed through several steps. The energy resolution of each ECAL module has been determined at a test beam~\cite{Arefev:2008zza} to be in agreement with the designed value of $\frac{\sigma_{\mathrm{E}}}{\mathrm{E}~(\mathrm{GeV)}} = \frac{10\%}{\sqrt{\mathrm{E}}}\oplus1\%$~\cite{Amato:494264}. The photoelectron multipliers gains are determined with the help of an LED system installed in the calorimeter. The final step of the calibration procedure is performed with several methods using the data coming from the experimental setup. First the energy flow method is applied, which is based on the idea that transverse energy flow measured by the ECAL is a locally smooth function~\cite{Voronchev:984656}. Further the calibration with neutral pions is performed. This method consists in determining the neutral pion peak position in $\piz\to\g\g$ decays~\cite{Puig:1348434, Perret:2014owa}. 

The performance of high energy photon reconstruction is illustrated by the reconstructed \mbox{$\mathrm{B}^{0}\to\mathrm{K}^{*0}\g$} and \mbox{$\mathrm{B}^{0}\to\upphi\g$} mass distribution~\cite{LHCb:2012ab}. The mass resolution obtained for this radiative decay, ∼92~\mevcc, is dominated by the ECAL energy resolution. Neutral pions with low transverse energy are mostly reconstructed as a resolved pair of well separated photons. A mass resolution of~8~\mevcc is obtained for such neutral pions.

In 2011(2012) LHCb was taking data in proton-proton collisions at a centre-of-mass energy of $\sqrt{\mathrm{s}}=7~\mathrm{TeV}$ ($8~\mathrm{TeV}$).  The collected data samples correspond to an integrated luminosity of~$1~\mathrm{fb}^{-1}$ in 2011 and $2~\mathrm{fb}^{-1}$ in 2012. The average number of visible pp interactions per bunch crossing, $\mu$, is 1.5 for both periods, which is four times larger than the designed value. The instantaneous luminosity, $\mathcal{L}$, was increased during 2011 up to~$4\times10^{-32}~\mathrm{cm}^{-2}\mathrm{c}^{-1}$ and during 2012 was kept constant at this value which is two times larger than the designed value. The excess of $\mu$ and $\mathcal{L}$ in comparison with the design values of $2\times10^{-32}~\mathrm{cm}^{-2}\mathrm{c}^{-1}$ leads to higher event multiplicities, which also implies larger number of tracks and higher number of clusters in the electromagnetic calorimeter and, as consequence, higher number of overlapping clusters. Due to the difference in conditions of the data taking, the correction factors are calculated separately for 2011 and 2012 periods.

\section{Event selection}
\label{seq:evsel} 
The decays \mbox{$\Bu\to\jpsi{}\mathrm{K}^{*+}$} followed by \mbox{$\mathrm{K}^{*+}\to\mathrm{K}^+\piz$} and \mbox{$\Bu\to\jpsi\mathrm{K}^{+}$} are reconstructed using the \mbox{$\jpsi\to\mumu$} decay mode. The signal candidates are selected with the help of cuts applied to various kinematic parameters, particle identification and decay reconstruction quality variables. Most selection criteria are common for the decay channels, except those which are related to the photons selection.

The track quality of reconstructed charged particles is ensured by requiring the $\chi^2$ per degree of freedom, $\chi^2_{\rm{tr}}/\mathrm{ndf}$, provided by the track fit to be less than three. Well-identified muons are selected by requiring that the difference of logarithms of the muon and hadron hypothesis likelihood, provided by the particle identification detectors~\cite{Powell:2010zz}, $\mathcal{L}_{\mu} > \mathcal{L}_{\pi}$, is larger than~zero.

Pairs of oppositely-charged tracks identified as muons, each having a transverse momentum greater than~$550~\mathrm{MeV}/c$ are combined to form \mbox{$\jpsi\to\mumu$} candidates. The invariant mass of dimuon combination is required to be between~3050\mevcc~and~3150\mevcc for \jpsi~candidates. Only events explicitly triggered by the selected \jpsi~candidates at all trigger levels have been considered in the analysis.

Photons are reconstructed using the electromagnetic calorimeter~\cite{LHCb-2003-091,LHCb-2003-092} and identified using a likelihood-based estimator, constructed from variables that rely on calorimeter and tracking information. Candidate photon clusters must not be matched to the~trajectory of a~track extrapolated from the~tracking system to the~cluster position in the~calorimeter. Further photon quality refinement is done using information from the~\presh and \spd detectors. The photon transverse momentum is required to be greater than~$250~\mevc$. 

For the photon identification a variable is used, which combines information from the calorimeter and tracking systems in order to reduce the background from hadrons, electrons and merged \piz (those which daughter photons create largely overlapping clusters in the ECAL). The value of this variable, CL, is required to be more than~2\%. The invariant mass of the photon pair is required to be within~30~\mevcc around the nominal~\piz~mass.

Charged kaons are selected by requiring the transverse momentum to be in excess of~$450~\mathrm{MeV}/c$. Loose kaon identification criterion is applied, the corresponding neutral network variable, $\mathcal{P}_{\mathrm{K}}$, is required to be greater than~0.1. The invariant mass of the $\mathrm{K^+} \piz$~combinations is required to be in range~$\pm~75\mevcc$~around the nominal $\mathrm{K}^{*+}$~mass~\cite{PDG-2014}.

The \Bu~candidates are formed from \mbox{$\jpsi\mathrm{K}^{\left(*\right)+}$}~pairs. A refit~\cite{Hulsbergen:2005pu} of the~\Bu~candidate is performed taking into account primary vertex pointing constraint and \jpsi~and \piz~mass constraints. The reduced $\chisq$ for this procedure, $\chisq_{\mathrm{fit}}/\mathrm{ndf}$, is required to be less than five. The transverse momentum of the \Bu~candidate is required to be larger than~$3~\gevc$. To reject possible contributions to \mbox{$\Bu\to\jpsi\mathrm{K}^{*+}$}~decay from \mbox{$\Bu\to\jpsi\mathrm{K}^{+}$}~decays with two additional random soft photons, the invariant mass of the \mbox{$\jpsi{}\mathrm{K}^+$}~combination is required to be outside a~$\pm25~\mevcc$~mass window around the known \Bu~mass. Finally the decay time, $c\tau$, of \Bu~candidates is required to be in excess of~$200~\mum$.

\section{Signal yields}
\label{sec:signal}

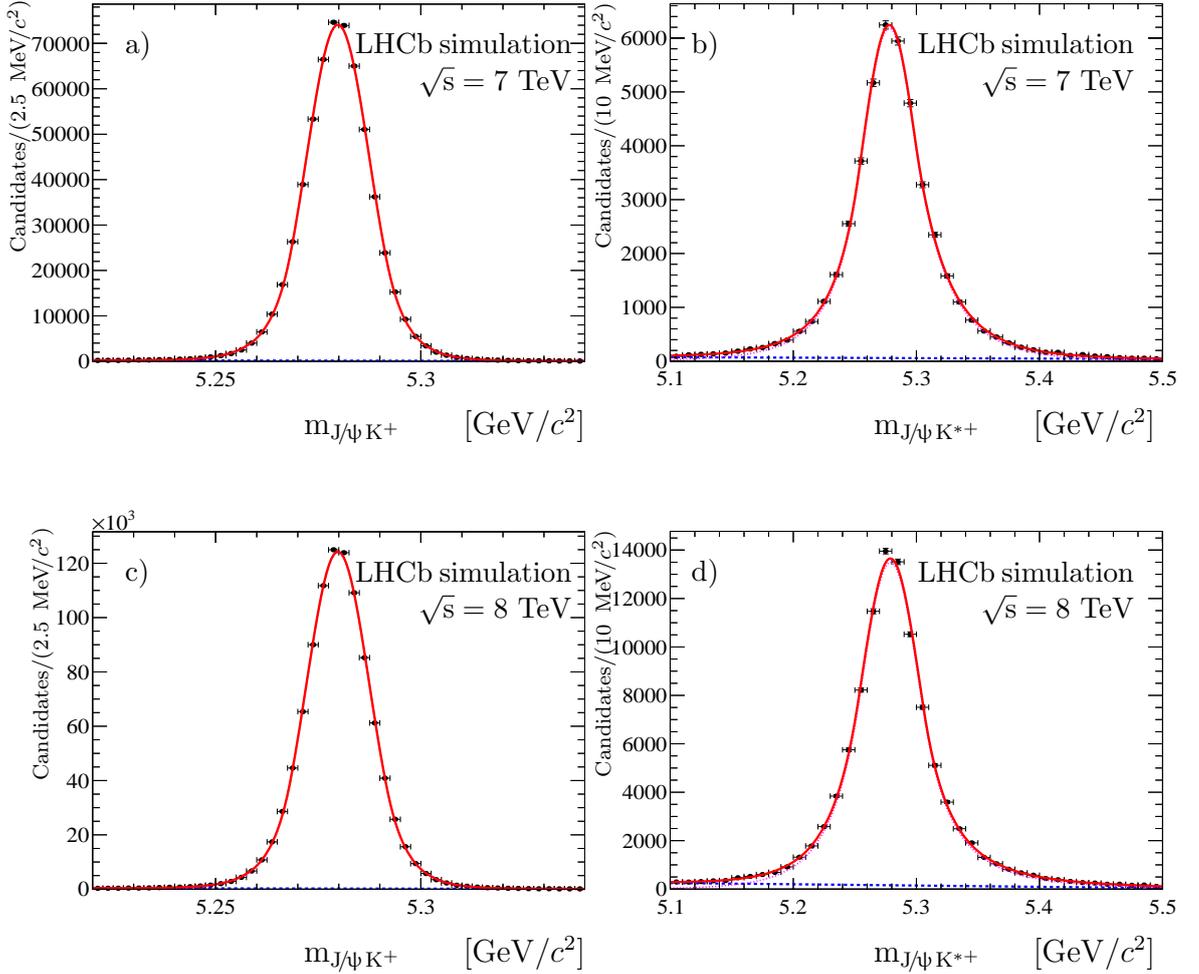
\begin{figure}[htb]
  \setlength{\unitlength}{1mm}
  \centering
  \begin{picture}(150,130)
    \put(0,70){
      \includegraphics*[width=80mm,height=60mm%
      ]{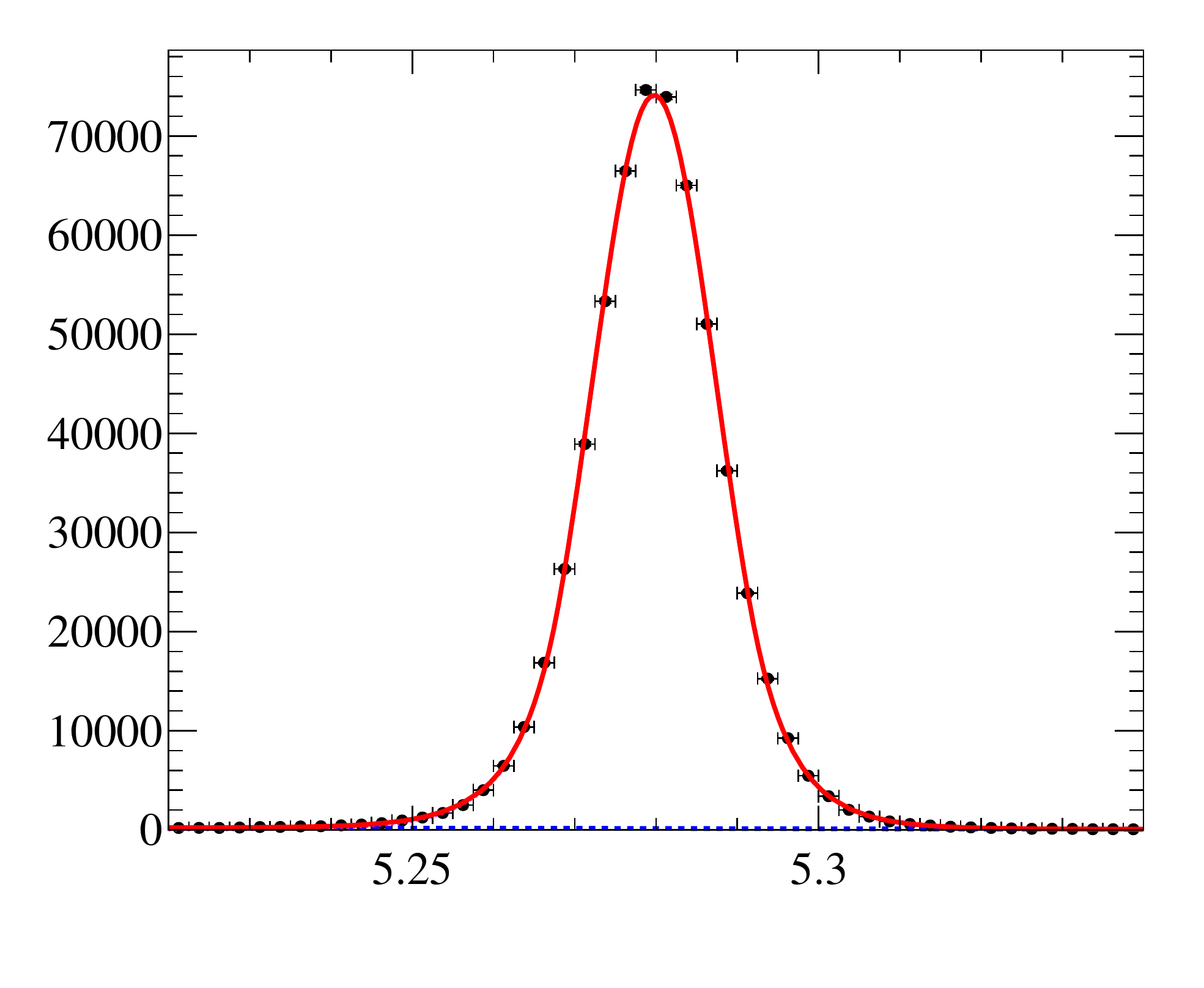}
    }
    \put(76,70){
      \includegraphics*[width=80mm,height=60mm,%
      ]{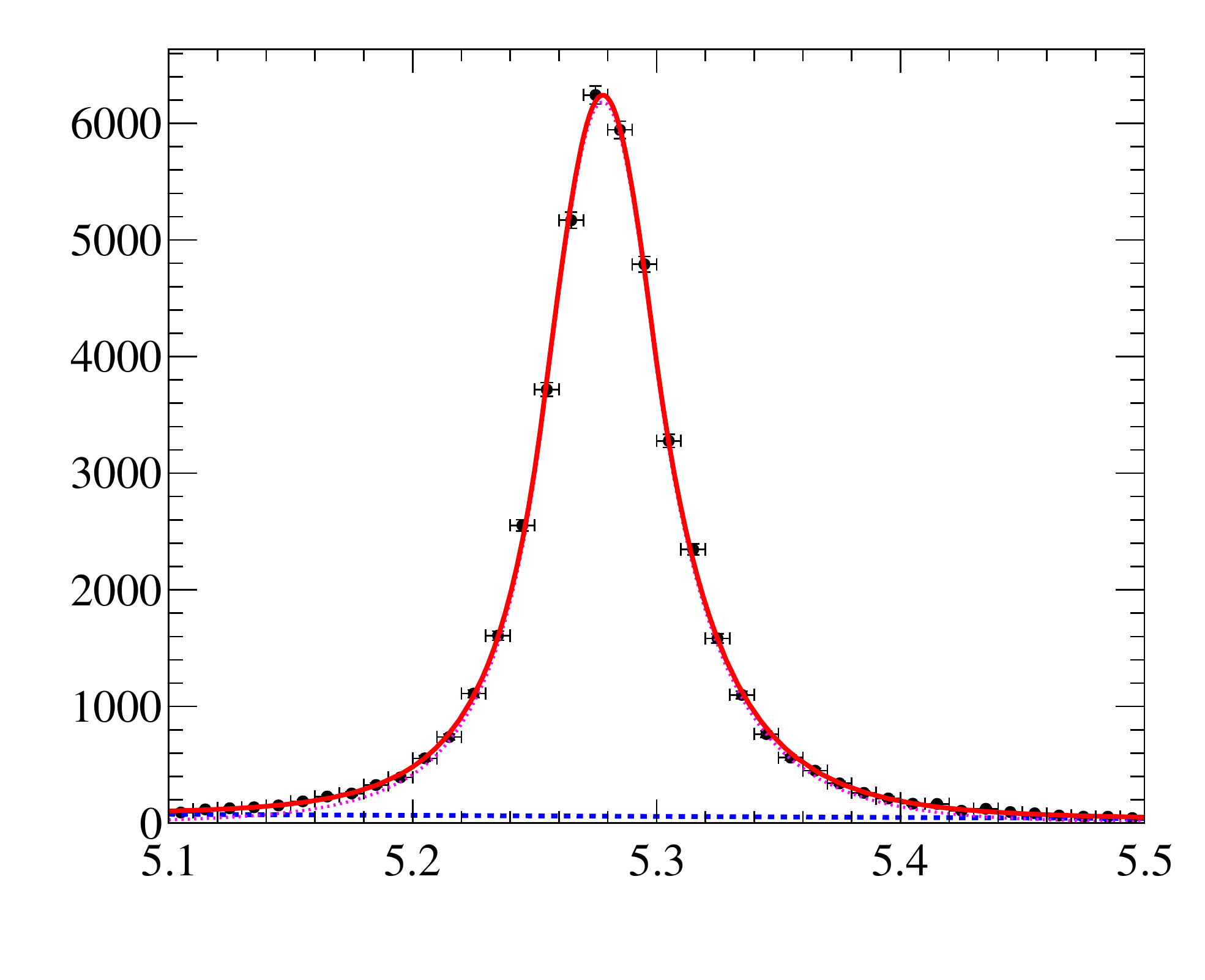}
    }
    \put(0,0){
      \includegraphics*[width=80mm,height=60mm%
      ]{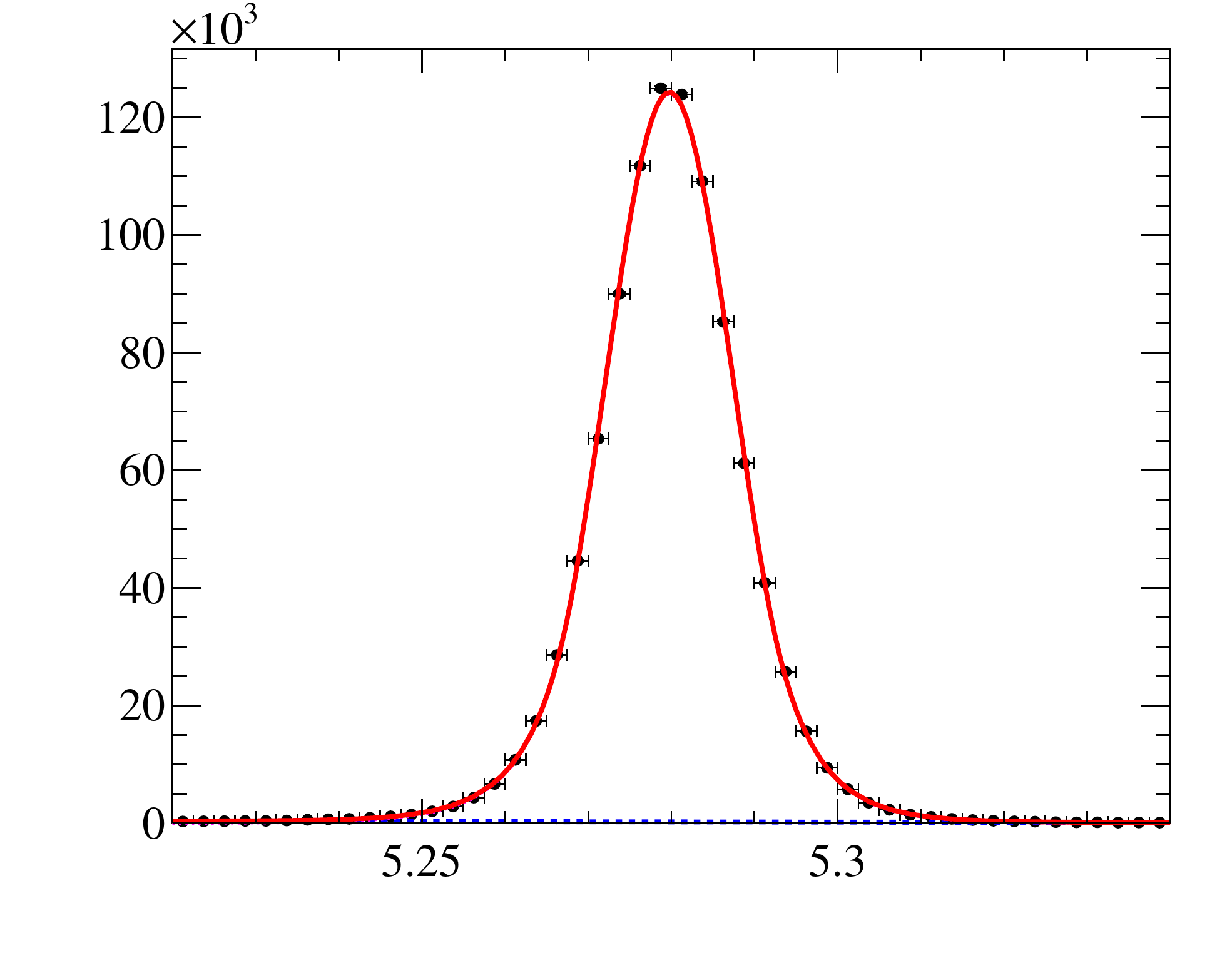}
    }
    \put(76,0){
      \includegraphics*[width=80mm,height=60mm,%
      ]{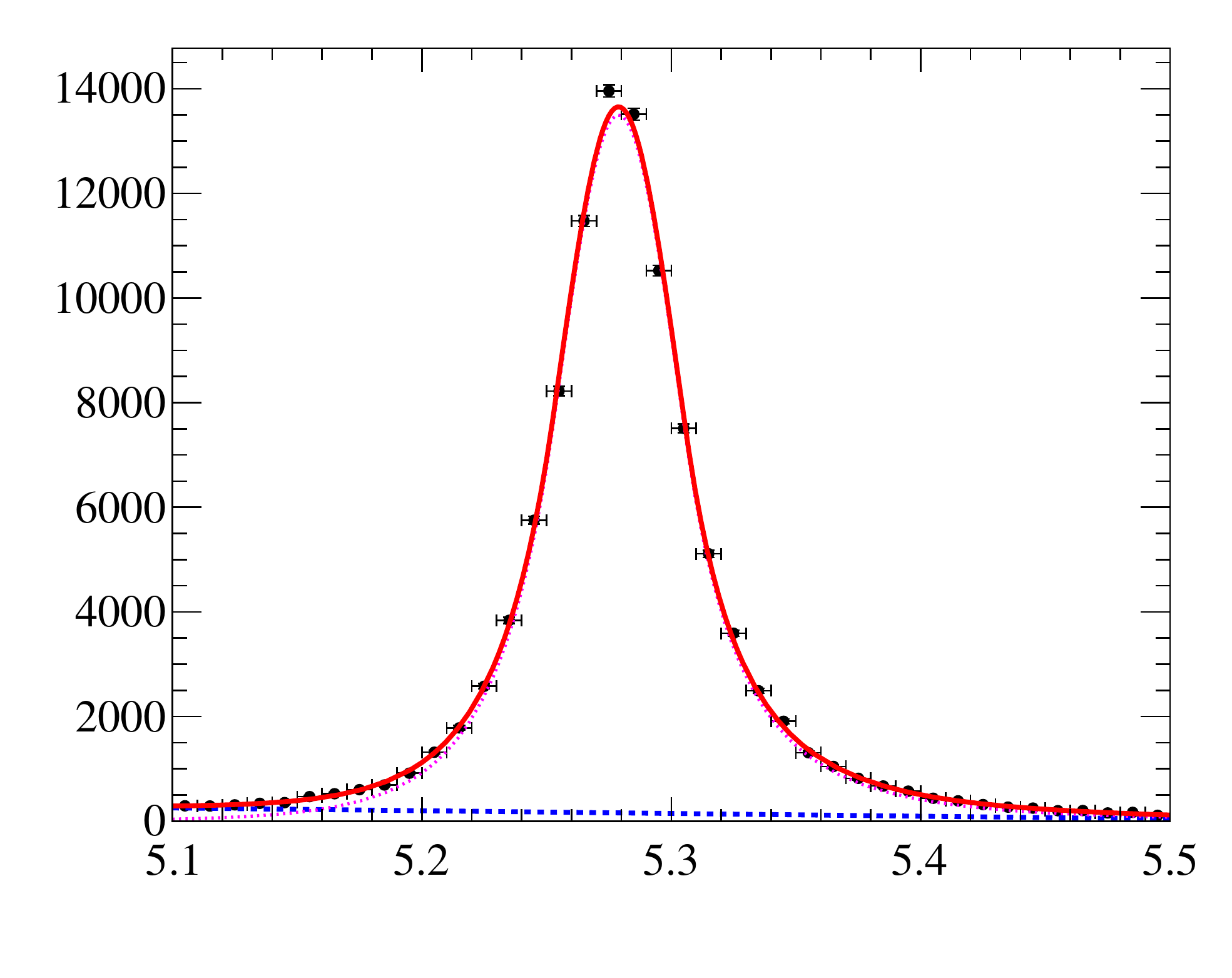}
    }

    \put( 39,70)  { $\mathrm{m}_{\jpsi\mathrm{K}^{+}}$     }
    \put(114,70)  { $\mathrm{m}_{\jpsi\mathrm{K}^{*+}}$    }
    \put( 60,70)  { $\left[ \mathrm{GeV}/c^2\right]$       }
    \put(135,70)  { $\left[ \mathrm{GeV}/c^2\right]$       }
    \put( 39, 0)  { $\mathrm{m}_{\jpsi\mathrm{K}^{+}}$     }
    \put(114, 0)  { $\mathrm{m}_{\jpsi\mathrm{K}^{*+}}$    }
    \put( 60, 0)  { $\left[ \mathrm{GeV}/c^2\right]$       }
    \put(135, 0)  { $\left[ \mathrm{GeV}/c^2\right]$       }
    \put(15.5 ,50.5) {  \small {c)}}
    \put(90,50.5)    {  \small {d)}}
    \put(15.5 ,120.5){  \small {a)}}
    \put(90,120.5)   {  \small {b)}}
    \put(44,48)  {   \small
      $\begin{array}{r}
        \mathrm{LHCb\medspace simulation} \\
        \sqrt{\mathrm{s}}=8~\mathrm{TeV}           \\
      \end{array}$
    }
    \put(118,48) {  \small
      $\begin{array}{r}
        \mathrm{LHCb\medspace simulation} \\
        \sqrt{\mathrm{s}}=8~\mathrm{TeV}           \\
      \end{array}$
    }
    \put(44,118) {   \small
      $\begin{array}{r}
        \mathrm{LHCb\medspace simulation} \\
        \sqrt{\mathrm{s}}=7~\mathrm{TeV}           \\
      \end{array}$
    }
    \put(118,118){  \small
      $\begin{array}{r}
        \mathrm{LHCb\medspace simulation} \\
        \sqrt{\mathrm{s}}=7~\mathrm{TeV}           \\
      \end{array}$
    }
    \put(77,24.5)  {  \scriptsize
    \begin{sideways}%
      Candidates/(10~\mevcc)
    \end{sideways}%
    } 
    \put(77,94.5)  {  \scriptsize
    \begin{sideways}%
      Candidates/(10~\mevcc)
    \end{sideways}%
    }
    \put(3,24)  {  \scriptsize
    \begin{sideways}%
      Candidates/(2.5~\mevcc)
    \end{sideways}%
    }
    \put(0,94)  {  \scriptsize
    \begin{sideways}%
      Candidates/(2.5~\mevcc)
    \end{sideways}%
    }
  \end{picture}
  \caption {
    Simulated mass distributions for the selected \mbox{$\Bu\to\jpsi\mathrm{K}^{+}$}~(left) and \mbox{$\Bu\to\jpsi\mathrm{K}^{*+}$}~(right) candidates (a,b) for 2011 and (c,d) 2012 data. Red curves represent the total fit function. Dashed purple and dashed blue curves represent signal and background respectively.
  }
  \label{fig:mcsignals}
\end{figure}

\begin{figure}[hb]
  \setlength{\unitlength}{1mm}
  \centering
  \begin{picture}(150,130)
    \put(1,70){
      \includegraphics*[width=75mm,height=60mm%
      ]{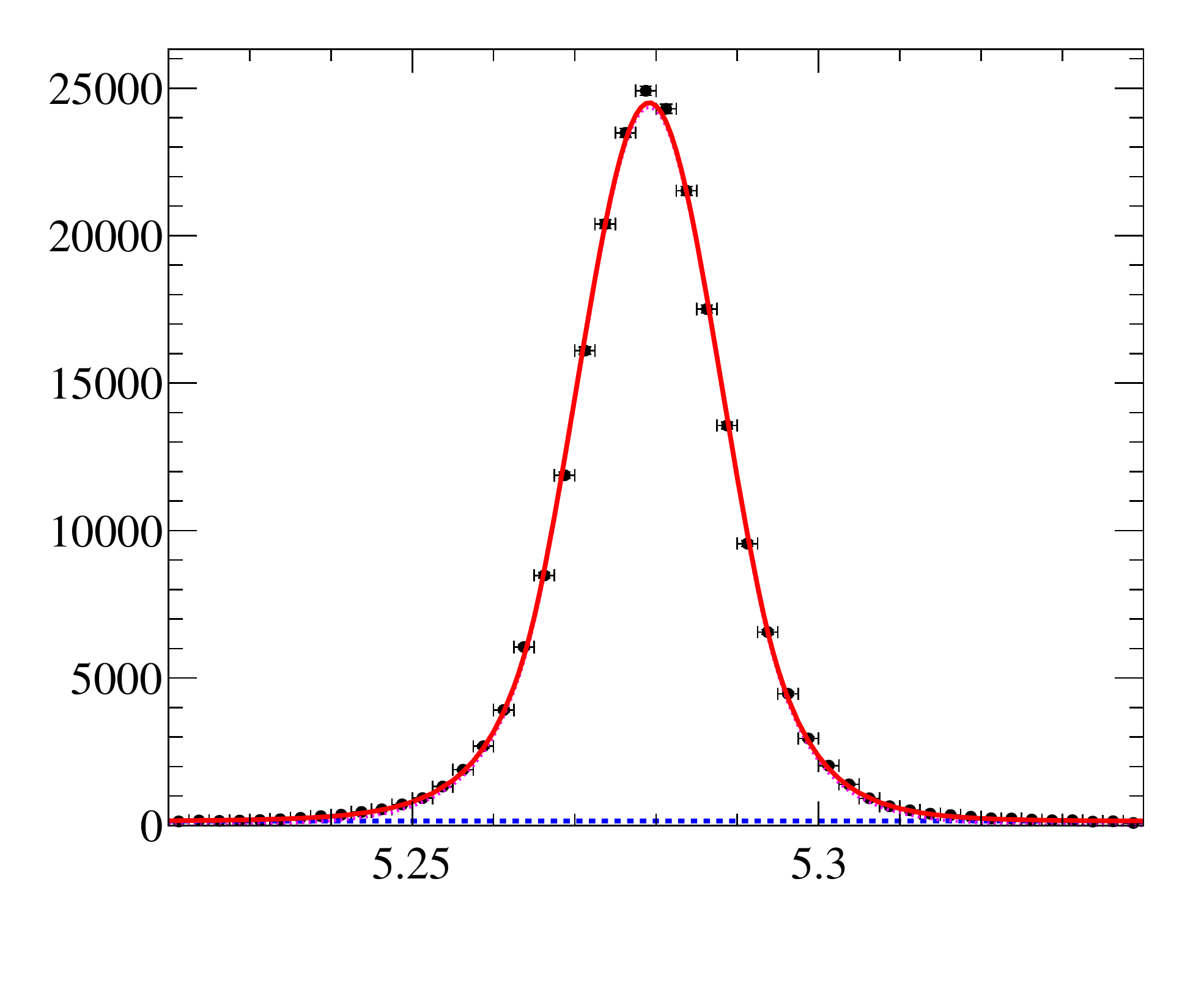}
    }
    \put(75,70){
      \includegraphics*[width=75mm,height=60mm,%
      ]{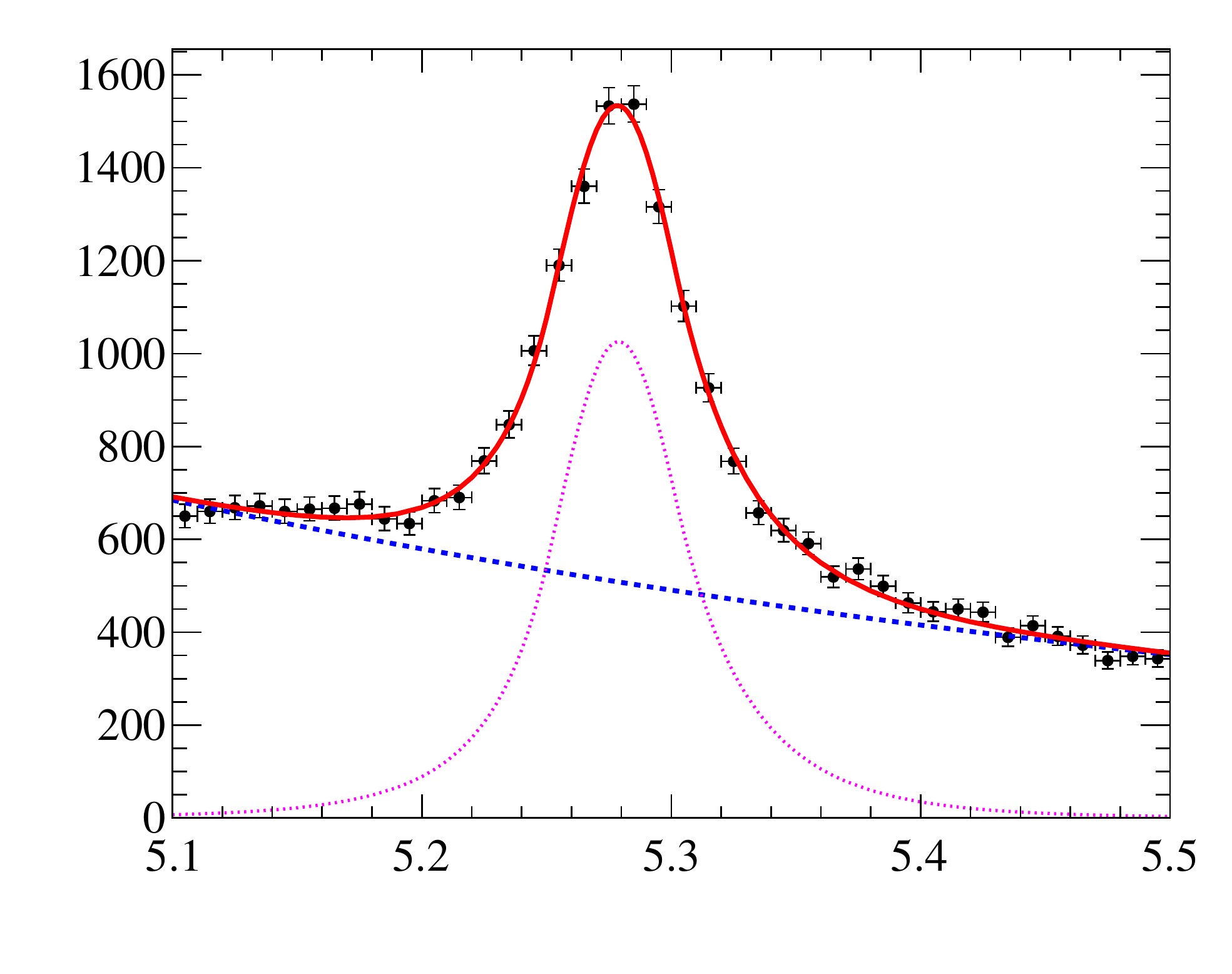}
    }
    \put(1,0){
      \includegraphics*[width=75mm,height=60mm%
      ]{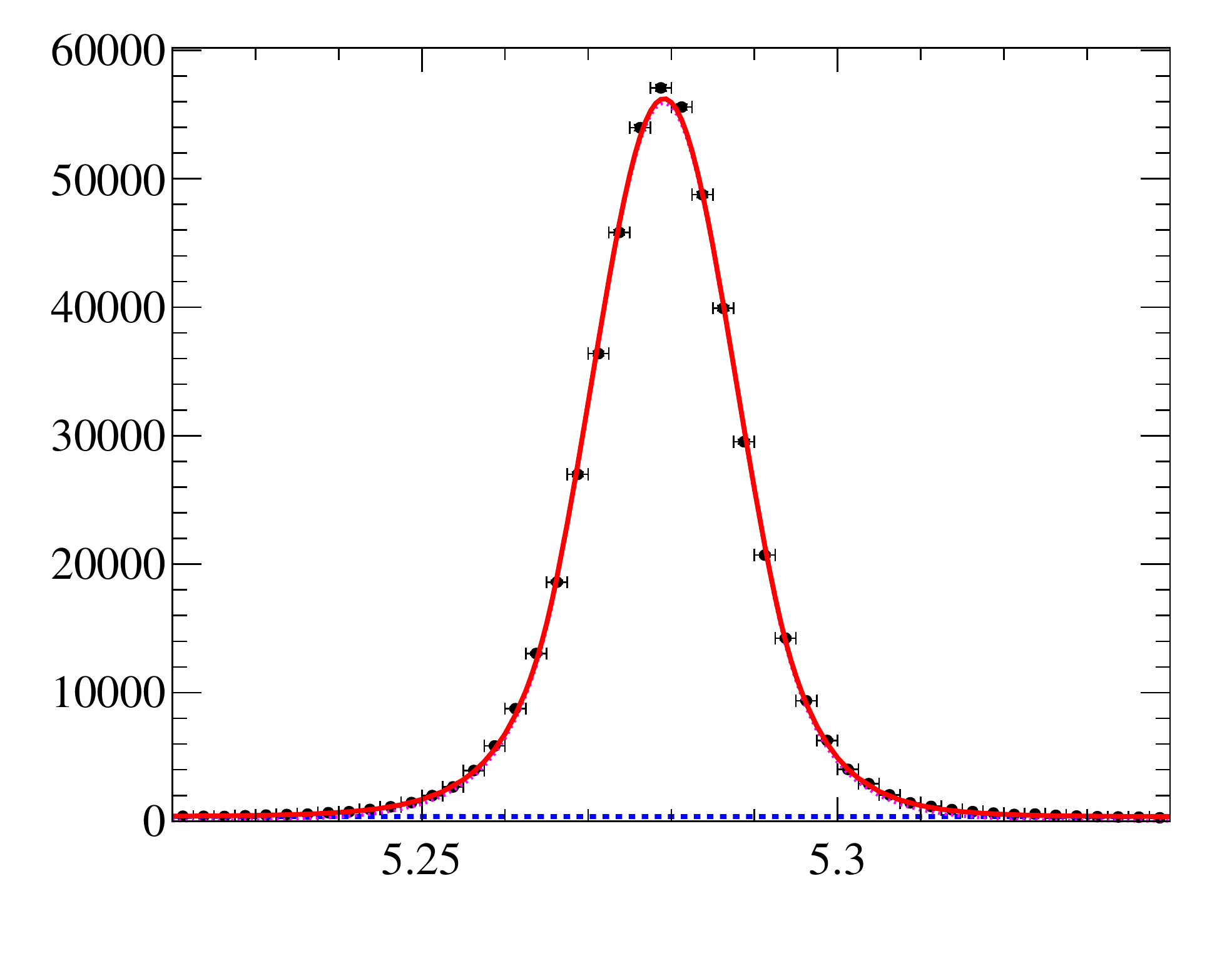}
    }
    \put(75,0){
      \includegraphics*[width=75mm,height=60mm,%
      ]{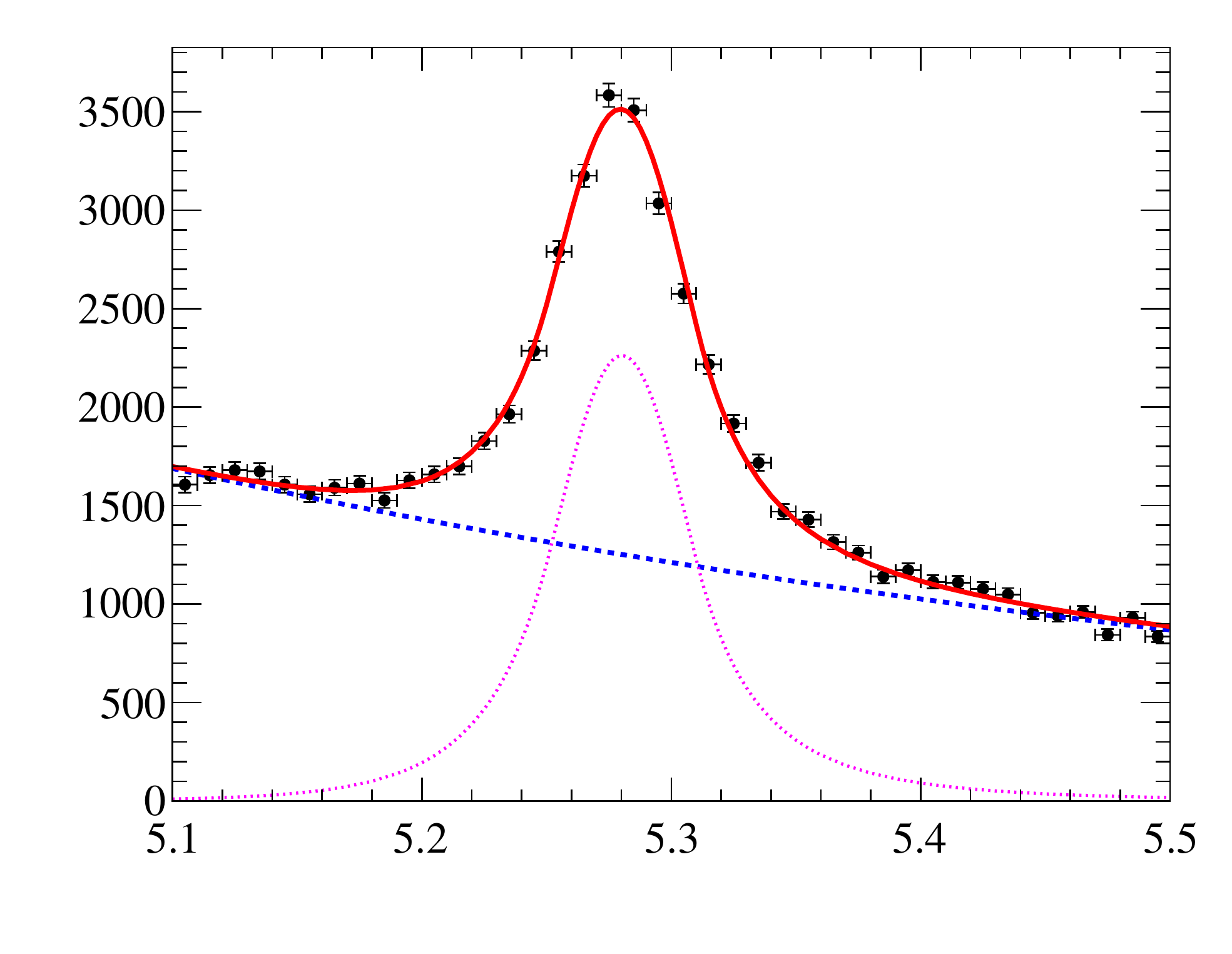}
    }

    \put( 35,70)  { $\mathrm{m}_{\jpsi\mathrm{K}^{+}}$     }
    \put(110,70)  { $\mathrm{m}_{\jpsi\mathrm{K}^{*+}}$    }
    \put( 56,70)  { $\left[ \mathrm{GeV}/c^2\right]$       }
    \put(131,70)  { $\left[ \mathrm{GeV}/c^2\right]$       }
    \put( 35, 0)  { $\mathrm{m}_{\jpsi\mathrm{K}^{+}}$     }
    \put(110, 0)  { $\mathrm{m}_{\jpsi\mathrm{K}^{*+}}$    }
    \put( 56, 0)  { $\left[ \mathrm{GeV}/c^2\right]$       }
    \put(131, 0)  { $\left[ \mathrm{GeV}/c^2\right]$       }
    \put(53 ,50.5) {  \small {c)}}
    \put(126,50.5) {  \small {d)}}
    \put(53 ,120.5){  \small {a)}}
    \put(126,120.5){  \small {b)}}
    \put(46,48)  {   \small
      $\begin{array}{r}
        \mathrm{LHCb}                 \\
        \sqrt{\mathrm{s}}=8~\mathrm{TeV}           \\
      \end{array}$
    }
    \put(119,48) {  \small
      $\begin{array}{r}
        \mathrm{LHCb}                 \\
        \sqrt{\mathrm{s}}=8~\mathrm{TeV}           \\
      \end{array}$
    }
    \put(46,118) {   \small
      $\begin{array}{r}
        \mathrm{LHCb}                 \\
        \sqrt{\mathrm{s}}=7~\mathrm{TeV}           \\
      \end{array}$
    }
    \put(119,118){  \small
      $\begin{array}{r}
        \mathrm{LHCb}                 \\
        \sqrt{\mathrm{s}}=7~\mathrm{TeV}           \\
      \end{array}$
    }
    \put(75,24.5)  {  \scriptsize
    \begin{sideways}%
      Candidates/(10~\mevcc)
    \end{sideways}%
    } 
    \put(75,95)  {  \scriptsize
    \begin{sideways}%
      Candidates/(10~\mevcc)
    \end{sideways}%
    }
    \put(0,25.0)  {  \scriptsize
    \begin{sideways}%
      Candidates/(2.5~\mevcc)
    \end{sideways}%
    }
    \put(0,94)  {  \scriptsize
    \begin{sideways}%
      Candidates/(2.5~\mevcc)
    \end{sideways}%
    }
    
  \end{picture}
  \caption {
    Invariant mass distributions for the selected \mbox{$\Bu\to\jpsi\mathrm{K}^{+}$}~(left) and \mbox{$\Bu\to\jpsi\mathrm{K}^{*+}$}~(right) candidates (a,b) for 2011 and (c,d) 2012 data. Red curves represent the total fit function. Dashed purple and dashed blue curves represent signal and background respectively.
  }
  \label{fig:signals}
\end{figure}

\begin{figure}[h!tb]
  \setlength{\unitlength}{1mm}
  \centering
  \begin{picture}(150,60)
    \put(35,0){
      \includegraphics*[width=85mm,height=70mm,%
      ]{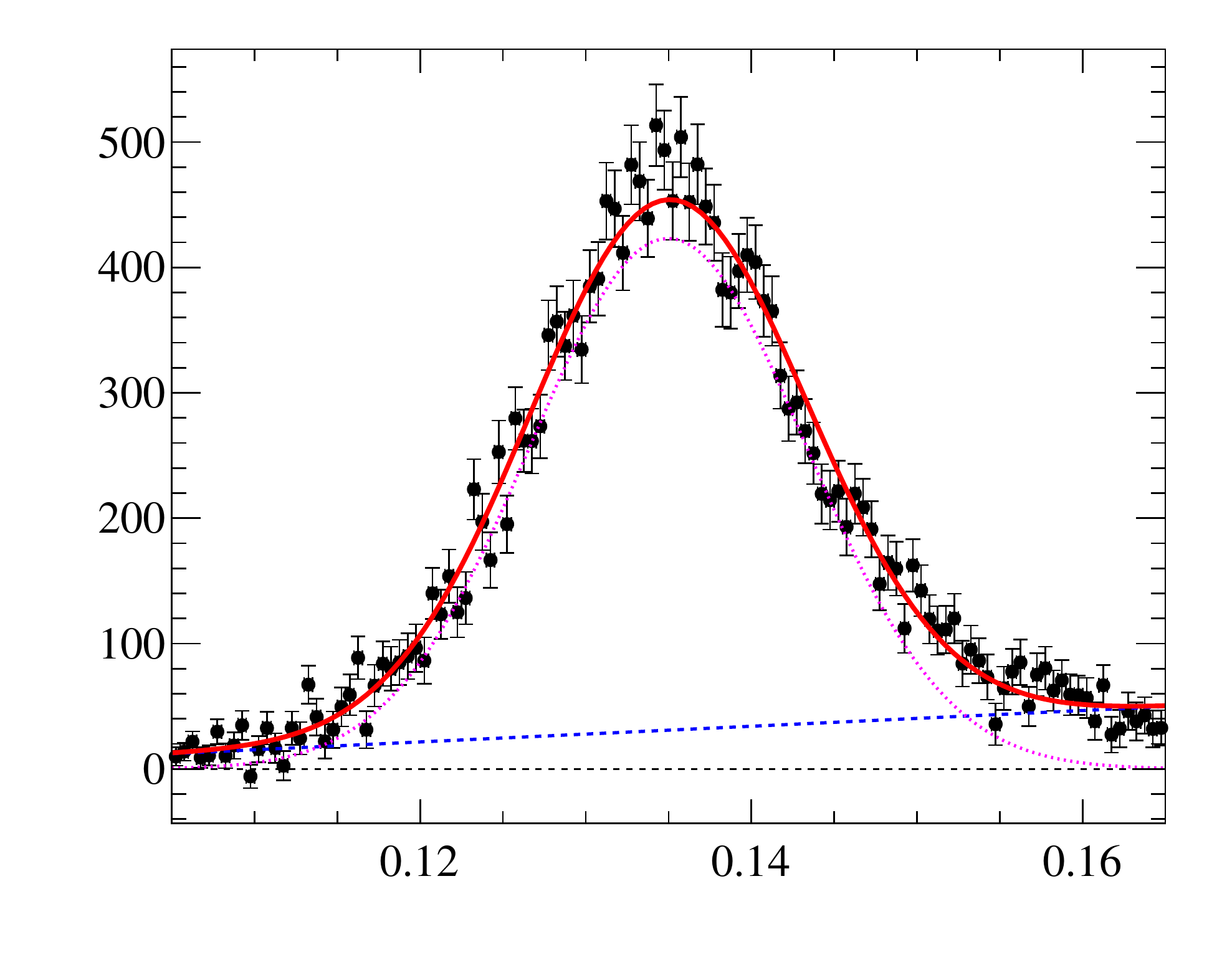}
    }
    \put(79,   0)  { $\mathrm{m}_{\g\g}$                  }
    \put(101,  0)  { $\left[ \mathrm{GeV}/c^2\right]$     }
    \put(91,55){  \small
      $\begin{array}{r}
        \mathrm{LHCb}                 \\
        \sqrt{\mathrm{s}}=8~\mathrm{TeV}           \\
      \end{array}$
    }
    \put(36,35)  {  \scriptsize
    \begin{sideways}%
      Candidates/(5~\mevcc)
    \end{sideways}%
    }
  \end{picture}
  \caption {
    Invariant mass distributions for the selected $\piz\to\gamma\gamma$ candidates.
  }
  \label{fig:pi0}
\end{figure}

The invariant mass distributions for the simulated \mbox{$\Bu\to\jpsi\mathrm{K}^{+}$} and \mbox{$\Bu\to\jpsi\mathrm{K}^{*+}$} events, which have passed the selection, are shown in Fig.~\ref{fig:mcsignals}. The mass distributions for the \mbox{$\Bu\to\jpsi\mathrm{K}^{+}$} and \mbox{$\Bu\to\jpsi\mathrm{K}^{*+}$} candidates selected from the data are shown in~Fig.~\ref{fig:signals}. The diphoton invariant mass distribution in data obtained using the sPlot technique~\cite{Pivk:2004ty} is shown in~Fig.~\ref{fig:pi0}. A clear peak from the $\piz\to\gamma\gamma$ decays is seen in the distribution. Its position is found to be $134.9 \pm 0.1 \mevcc$ and the resolution is $8.3 \pm 0.1 \mevcc$.

The signal yields are determined using an extended unbinned maximum likelihood fit. The signal component for both decays is modelled by a double-sided Crystal Ball function~\cite{Skwarnicki:1986xj}. The tail parameters of the signal function are fixed using simulated events, whereas the mean and resolution of the \Bu~candidate are allowed to vary in the fit. The combinatorial background is modelled by an exponential function. For both final states the fitted position of the~\Bu~peak is consistent with the known~\Bu~mass~\cite{PDG-2014} and the mass resolution is in agreement with the values observed in simulated samples. 

\section{Efficiencies}
\label{sec:effic}
The efficiencies and their ratios are estimated using simulation. For the simulation, pp collisions are generated using \pythia~6 and \pythia~8 ~\cite{Sjostrand:2006za} with a specific \lhcb configuration~\cite{LHCb-PROC-2010-056}.  Decays of hadronic particles are described by \evtgen~\cite{Lange:2001uf} in which final state radiation is generated using \photos~\cite{Golonka:2005pn}. The interaction of the generated particles with the detector and its response are implemented using the \geant toolkit~\cite{Allison:2006ve, *Agostinelli:2002hh} as described in Ref.~\cite{LHCb-PROC-2011-006}. The digitized output is passed through a full simulation of both the hardware and software trigger and then reconstructed and selected in the same way as the data.

Ratios of efficiencies between \mbox{$\Bu\to\jpsi{}\mathrm{K}^{*+}\left(\to\mathrm{K}^+\piz\right)$} and \mbox{$\Bu\to\jpsi\mathrm{K}^{+}$}~decay modes are calculated separately to obtain the overall efficiency ratio. The following efficiency ratio is obtained for 2011 data
\begin{align*}
\frac{\varepsilon^{\mathrm{total}}_{\Bu\to\jpsi\mathrm{K}^{+}}}{\varepsilon^{\mathrm{total}}_{\Bu\to\jpsi\mathrm{K}^{*+}}}&=13.32\pm0.18,
\end{align*}
and for 2012 data
\begin{align*}
\frac{\varepsilon^{\mathrm{total}}_{\Bu\to\jpsi\mathrm{K}^{+}}}{\varepsilon^{\mathrm{total}}_{\Bu\to\jpsi\mathrm{K}^{*+}}}&=13.57\pm0.12,
\end{align*}

\noindent where the uncertainty is due to the limited size of simulated samples. 

\section{Systematic uncertainties}
\label{seq:system}

Since the decay products in each channel have similar kinematic properties, most uncertainties cancel in the ratio, in particular, those related to the muon and $\jpsi$ reconstruction and identification. The remaining systematic uncertainties are discussed below.

Systematic uncertainties related to the fit model are estimated using alternative models for the description of the invariant mass distributions. The following functions are used as alternative:
\begin{itemize}
\item a sum of a Student-t distribution with all parameters fixed from the simulation for the signal and exponent for background;
\item a sum of modified Novosibirsk function~\cite{Lees:2011gw} with all parameters fixed from the simulation for the signal and exponent for background;
\item a sum of Crystall Ball function for the signal and second order polynomial for background.
\end{itemize}
For each fit model, the ratio of event yields is calculated. The maximum observed deviation from the baseline model in the ratio of yields is taken as systematic uncertainty. The obtained uncertainty is~0.2~$\%$.

The uncertainty due to the limited number of simulated events vary in a range from 0.8\% to 4.5\%, depending on the requirements on the photon transverse energy and neutral pion transverse momentum. The corresponding systematic uncertainty is assigned in each bin of the photon transverse momentum individually.

The trigger efficiencies for events with \mbox{$\jpsi\to\mumu$} produced in beauty hadron decays is studied and a systematic uncertainty~1.1~$\%$ is assigned based on the comparison of the ratio of trigger efficiencies for high-yield samples of $\Bu\to\jpsi\mathrm{K}^+$ and $\Bu\to \Ppsi(2\mathrm{S}) \mathrm{K}^+$~decays on data and simulation~\cite{Aaij:1484958}.

The acceptance efficiency is calculated separately for different magnet polarities and for different data taking periods. The obtained individual acceptance efficiencies for the \mbox{$\Bu\to\jpsi{}\left(\mathrm{K}^{*+}\to\mathrm{K}^+\piz\right)$} and \mbox{$\Bu\to\jpsi\mathrm{K}^{+}$}~decays and their ratios are represented in Table~\ref{table:Eff_acc}. The average of four ratios for~2011~(\mbox{$1.1040 \pm 0.0015$}) is chosen as the resulting ratio of acceptance efficiencies (uncertainty is statistical only). The maximum deviation from the average is taken as systematic uncertainty and equals~1.4~$\%$. The average of four ratios for 2012~(\mbox{$1.0998 \pm 0.0015$}) is chosen as the resulting ratio of acceptance efficiencies (uncertainty is statistical only). The maximum deviation from the average is taken as systematic uncertainty and equals~0.2~$\%$. The observed difference in the efficiency ratios is conservatively taken as an estimate of the systematic uncertainty and is~1.4~$\%$ for~2011 data and~0.2~$\%$ for~2012 data.

\begin{table}[htb]
\caption{Acceptance efficiencies with their ratios obtained 
  from simulation. Uncertainties on the efficiencies and their ratios are 
  statistical only and reflect the simulation statistics.}
\begin{center}
\begin{tabular}{c|c|c|c|c|c}
Period & Pythia & Magnet & \mbox{$\Bu\to\jpsi\mathrm{K}^{+}$} & \mbox{$\Bu\to\jpsi\mathrm{K}^{*+}$} & 
$\varepsilon^{\mathrm{gen\&acc}}_{\Bu\to\jpsi\mathrm{K}^{+}}$/$\varepsilon^{\mathrm{gen\&acc}}_{\Bu\to\jpsi\mathrm{K}^{*+}}$ \\
\hline
2011  & 6 & Up   & $0.1550 \pm 0.0003$ & $0.1396 \pm 0.0003$ & $1.111 \pm 0.003$ \\
      &   & Down & $0.1547 \pm 0.0003$ & $0.1401 \pm 0.0003$ & $1.104 \pm 0.003$ \\ 
      & 8 & Up   & $0.1636 \pm 0.0003$ & $0.1491 \pm 0.0003$ & $1.097 \pm 0.003$ \\
      &   & Down & $0.1640 \pm 0.0003$ & $0.1485 \pm 0.0003$ & $1.104 \pm 0.003$ \\ 
\hline
2012  & 6 & Up   & $0.1578 \pm 0.0003$ & $0.1432 \pm 0.0002$ & $1.102 \pm 0.003$ \\
      &   & Down & $0.1573 \pm 0.0003$ & $0.1431 \pm 0.0002$ & $1.099 \pm 0.003$ \\ 
      & 8 & Up   & $0.1665 \pm 0.0005$ & $0.1518 \pm 0.0002$ & $1.097 \pm 0.003$ \\
      &   & Down & $0.1675 \pm 0.0004$ & $0.1522 \pm 0.0003$ & $1.101 \pm 0.003$ \\ 
\end{tabular}
\end{center}
\label{table:Eff_acc}
\end{table} 

The agreement between data and simulation has also been cross-checked by varying the selection criteria. No unexpectedly large deviation is found and no contribution is taken as systematics.

The contribution to \mbox{$\Bu\to\jpsi\mathrm{K}^{+}$} signal yield from \mbox{$\Bu\to\jpsi\pip$}~decay mode, when pion is misidentified as kaon, is found to be negligible~(lower than~0.001\%). And the contribution to \mbox{$\Bu\to\jpsi\mathrm{K}^{*+}(\to\mathrm{K}^+\piz)$} signal yield from \mbox{$\Bu\to\jpsi\Prho^{+}(\to\pip\piz)$}~decay mode, when pion is misidentified as kaon, is also find to be negligible~(around~0.02\%). 

The summary of systematic uncertainties is presented in Table~\ref{table:syst}. The total systematic uncertainty is~$1.8\%$ for~2011 and~$1.1\%$ for~2012.

\begin{table}[htb]
  \centering
  \caption{ \small
    Relative systematic uncertainties
    for the correction factor of the reconstruction efficiency. The total uncertainty is the quadratic sum of the individual components.} 
  \label{table:syst}
  \vspace*{3mm}
  \begin{tabular*}{0.55\textwidth}{@{\hspace{1mm}}lc@{\extracolsep{\fill}}cc@{\hspace{2mm}}}
    \multirow{3}{*}{Source} & \multicolumn{2}{c}{Uncertainty~[\%]} \\ 
                                      & 2011  & 2012 \\     
                                      & \mbox{$\sqrt{s} = 7$~TeV} & \mbox{$\sqrt{s} = 8$~TeV} \\
    \hline
    Fit model                         &  \multicolumn{2}{c}{0.2}  
    \\
    Trigger                           &  \multicolumn{2}{c}{1.1}
    \\
    Acceptance                        &  1.4  &  0.2 
    \\
    \hline
    Total \medspace\medspace\medspace\medspace       &  1.8 & 1.1 
  \end{tabular*}
\end{table}

\section{Correction factor}
\label{sec:corr}

The correction factor is calculated according to Eq.~\ref{eq:main}. For 2011 data its value is calculated to be
\begin{equation*}
 \Peta^{\mathrm{corr}}_{\piz} = 
\left( 103.2 \pm 2.6\stat\pm 2.3\syst\pm 6.7(\BR) \right)~\%, 
\end{equation*}
and for 2012 data is calculated to be
\begin{equation*}
 \Peta^{\mathrm{corr}}_{\piz} = 
\left( 105.9 \pm 1.8\stat\pm 1.6\syst\pm 6.9(\BR) \right)~\%,
\end{equation*}
where the third uncertainty is related to an uncertainty of the known branching fractions of \mbox{$\Bu\to\jpsi\mathrm{K}^{*+}$}~and~\mbox{$\Bu\to\jpsi\mathrm{K}^{+}$}~decays~\cite{PDG-2014}.

Assuming that photon reconstruction efficiencies are uncorrelated the correction factor $\Peta^{\mathrm{corr}}_{\piz}$ is interpreted as the square of the photon reconstruction uncertainty $\Peta^{\mathrm{corr}}_{\piz} = \left( \Peta^{\mathrm{corr}}_{\g} \right)^2$. This leads to following values of the corrections for the transverse energy of the photon more than 250\mev. 
\begin{equation*}
 \Peta^{\mathrm{corr}}_{\g} = 
\left( 101.6 \pm 1.3\stat\pm 1.1\syst\pm 3.3(\BR) \right)~\%,
\label{eq:gamma_nc}
\end{equation*}
for 2011 data and for 2012 data is
\begin{equation*}
 \Peta^{\mathrm{corr}}_{\g} = 
\left( 102.9 \pm 0.9\stat\pm 0.8\syst\pm 3.4(\BR) \right)~\%, 
\label{eq:gamma_mc}
\end{equation*}
where the third uncertainty is related to an uncertainty of the known branching fractions of \mbox{$\Bu\to\jpsi\mathrm{K}^{*+}$}~and~\mbox{\mbox{$\Bu\to\jpsi\mathrm{K}^{+}$}}~decays.

\subsection{Correction factor for different $\mathrm{E}^{\mathrm{T}}(\g)$}
\label{sec:corr_for_gamma}

The correction factor $\Peta^{\mathrm{corr}}_{\g}$~is also studied in four bins of the photon transverse energy. The correction factors $\Peta^{\mathrm{corr}}_{\piz}$ can be represented as product of $\Peta^{\mathrm{corr}}_{\g}$ for first and second photons. This leads to ten correction factors for $\piz$, depending on the combination of the bins, hit by the daughter photons of the $\piz$. The system of equations is obtained and then solved using the $\chi^2$ method. During this procedure only statistical uncertainties are taken into account.

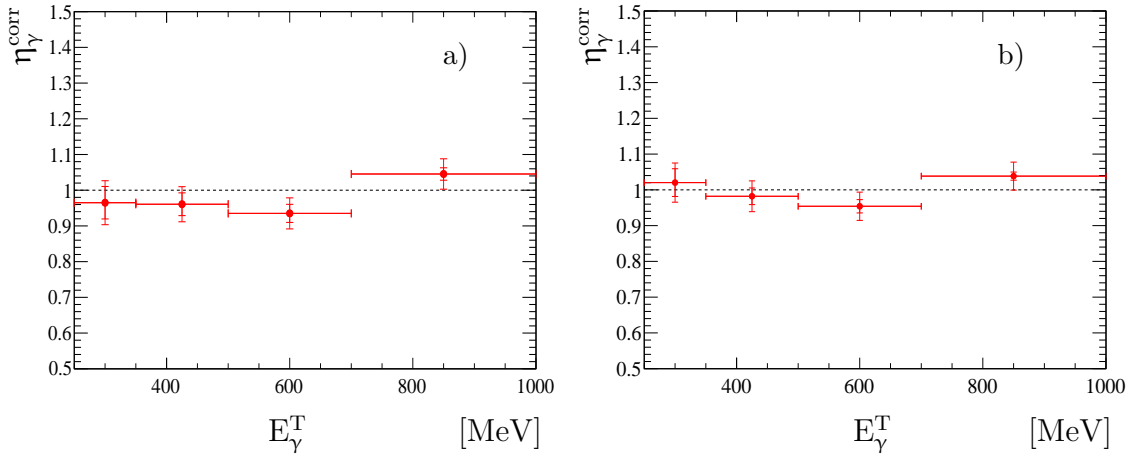
\begin{figure}[ht!b]
  \setlength{\unitlength}{1mm}
  \centering
  \begin{picture}(150,60)
    \put(0,0){
      \includegraphics*[width=75mm,height=60mm%
      ]{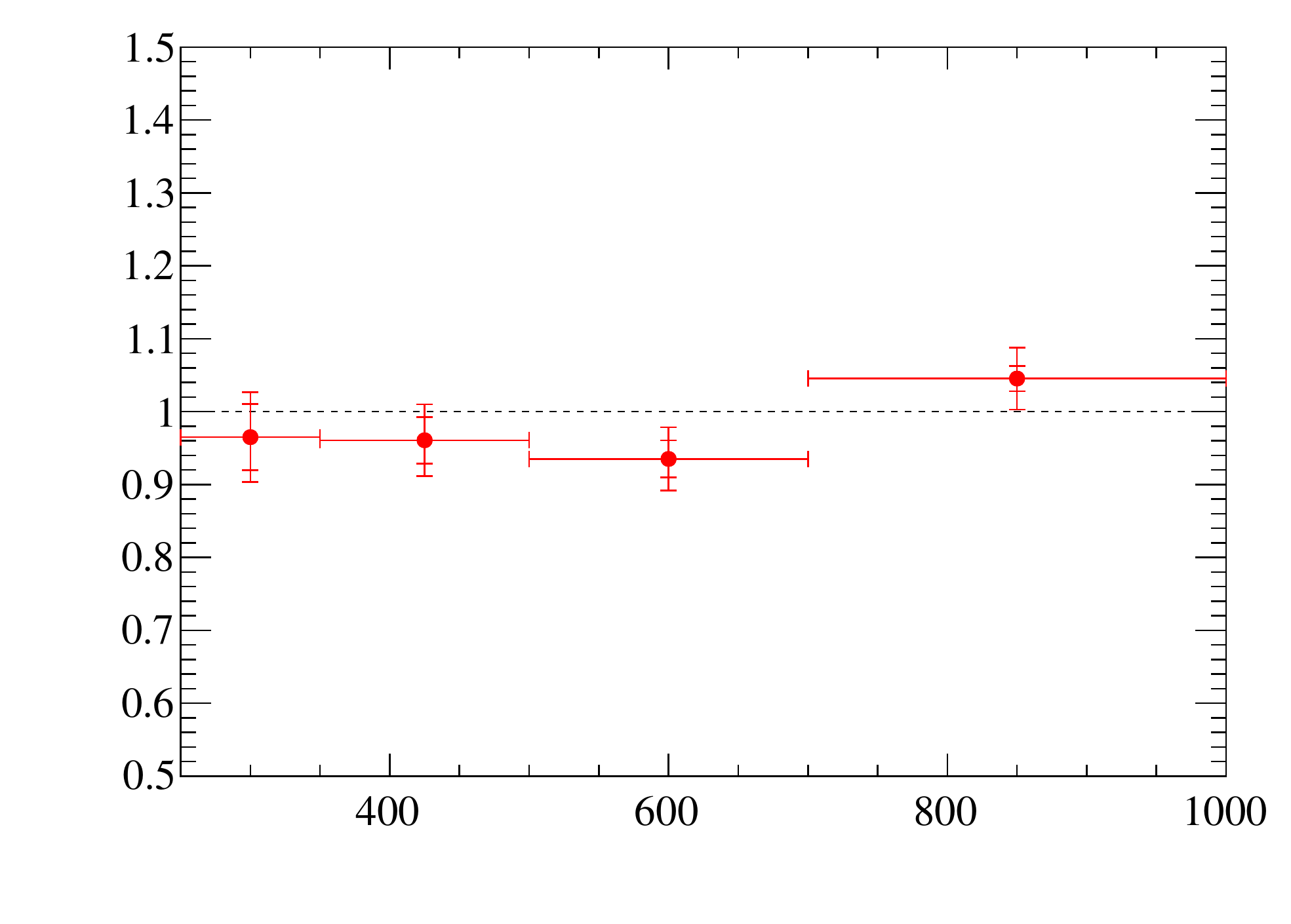}
    }
    \put(75,0){
      \includegraphics*[width=75mm,height=60mm,%
      ]{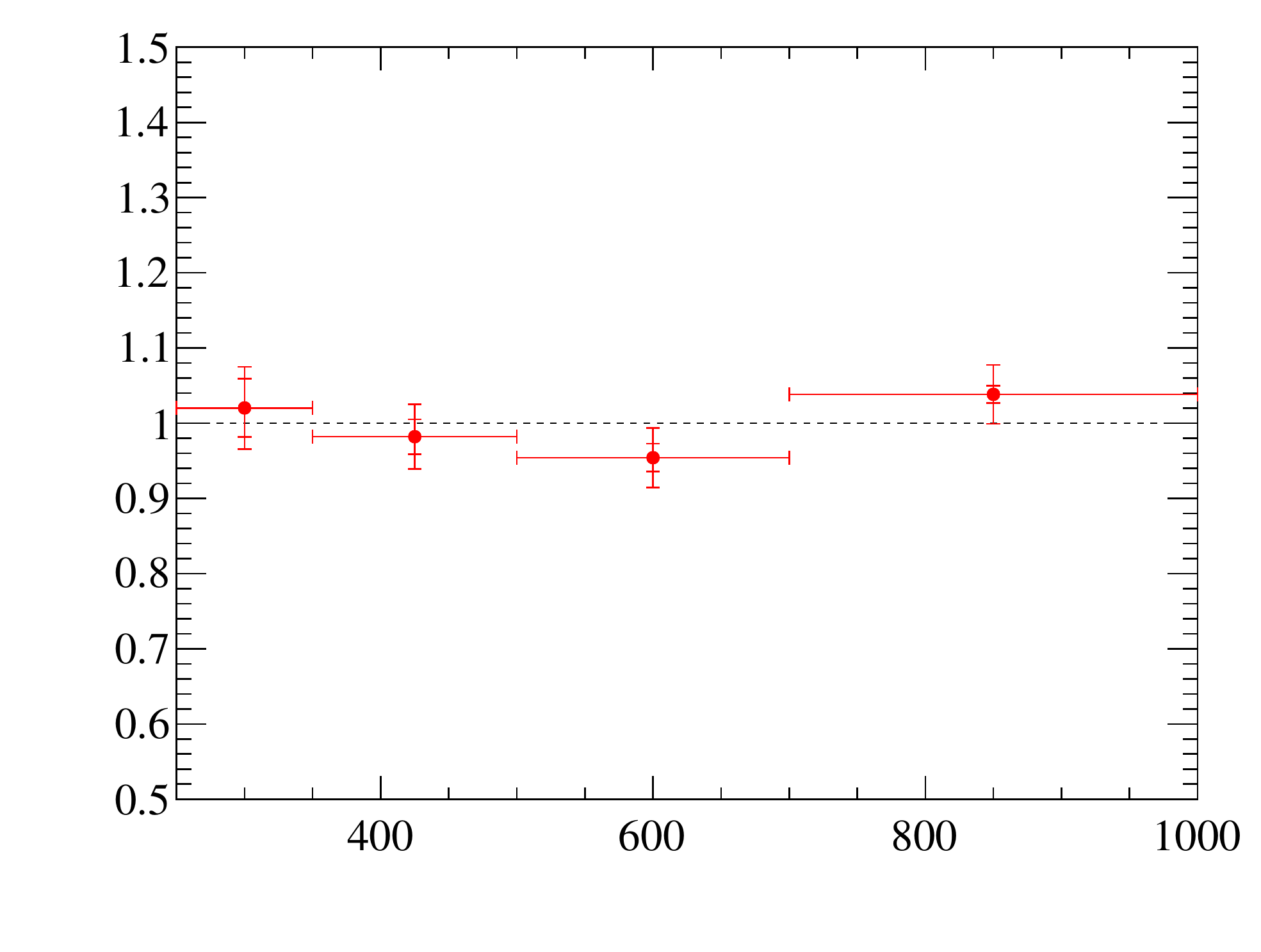}
    }
    \put( 36, 0)  { $\mathrm{E}^{\mathrm{T}}_{\g}$     }
    \put(113, 0)  { $\mathrm{E}^{\mathrm{T}}_{\g}$     }
    \put( 61, 0)  { $\left[ \mathrm{MeV}\right]$     }
    \put(136, 0)  { $\left[ \mathrm{MeV}\right]$     }
    \put(59 ,50) {  \small {a)}}
    \put(132,50) {  \small {b)}}

    \put(77,50) {
    \begin{sideways}%
      $\Peta^{\mathrm{corr}}_{\g}$
    \end{sideways}%
    } 
    \put(2, 50) {
    \begin{sideways}%
      $\Peta^{\mathrm{corr}}_{\g}$
    \end{sideways}%
    }
    
  \end{picture}
  \caption {
    Factors $\Peta^{\mathrm{corr}}_{\g}$~for (a) 2011 and (b) 2012  data in bins of photon transverse energy.
  }
  \label{fig:bin_gamma}
\end{figure}

In order to estimate the systematic uncertainties of the photon corrections, the corrections for the $\piz$ reconstruction efficiencies are varied within their systematic uncertainties. Each time the set of photon corrections is recalculated with the help of the $\chi^2$ method described above. In total 400 such experiments are performed. The 68\% interval of the variation of the photon reconstruction efficiency correction in each bin is taken as the systematic uncertainty for this bin.
The obtained values for $\Peta^{\mathrm{corr}}_{\g}$ are shown on Fig.~\ref{fig:bin_gamma} and certain values are listed in Table~\ref{table:pt_gamma}.

\begin{table}[htb!]
  \centering
  \caption{
    Factors $\Peta^{\mathrm{corr}}_{\g}$~for 2011 and 2012 data in bins of photon transverse energy. The first uncertainty is the statistical uncertainty, the second is systematic.
  } \label{table:pt_gamma}
  \vspace*{3mm}
  \begin{tabular*}{0.8\textwidth}{@{\hspace{1.5mm}}c@{\extracolsep{\fill}}cc@{\hspace{1.5mm}}}
    \hline
    $\mathrm{E}^{\mathrm{T}}_{\g}~~\left[\gevc\right]$
    & $\Peta^{\mathrm{corr}}_{\g}~~\left[\%\right]$ 2011 data
    & $\Peta^{\mathrm{corr}}_{\g}~~\left[\%\right]$ 2012 data
    \\
    \hline 
    250 -- 350
    & $\medspace\medspace\medspace\medspace\medspace\medspace  96.5 \pm 4.5 \pm 4.1$
    & $\medspace\medspace\medspace\medspace 102.0 \pm 3.9 \pm 3.5$
    \\
    350 -- 500  
    & $\medspace\medspace\medspace\medspace\medspace\medspace 96.1 \pm 3.2 \pm 3.4$ 
    & $\medspace\medspace\medspace\medspace\medspace\medspace 98.2 \pm 2.3 \pm 3.3$ 
    \\
    500 -- 700
    & $\medspace\medspace\medspace\medspace\medspace\medspace 93.5 \pm 2.5 \pm 3.3$ 
    & $\medspace\medspace\medspace\medspace\medspace\medspace 95.4 \pm 1.9 \pm 3.2$ 
    \\
    $\medspace\medspace\medspace\medspace\medspace\medspace\medspace > 700 $ 
    & $\medspace\medspace\medspace\medspace 104.5 \pm 1.7 \pm 3.6$
    & $\medspace\medspace\medspace\medspace 103.8 \pm 1.1 \pm 3.4$
    \\ 
    \hline 
  \end{tabular*}
\end{table}

\subsection{Correction factor for different $\mathrm{p}^{\mathrm{T}}(\piz)$}
\label{sec:corr_for_pi}

The dependence of the correction factor $\Peta^{\mathrm{corr}}_{\piz}$~on the neutral pion transverse momentum is also studied. The transverse momentum spectra are divided into a few bins and the analysis is repeated for each bin separately. The widths of the bins are varied to allow for sufficient number of candidates in each bin. The obtained values for $\Peta^{\mathrm{corr}}_{\piz}$ are listed in Table~\ref{table:pt_pi11} for 2011 data and in Tables~\ref{table:pt_pi12} for 2012 data. 

\begin{table}[h!tb]
  \centering
  \caption{
    Factors~$\Peta^{\mathrm{corr}}_{\piz}$~for 2011 data
    in bins of neutral pion transverse momentum. The first uncertainty is the statistical uncertainty,
    the second is systematic, and the third is related to an uncertainty
    of the known branching fractions of \mbox{$\Bu\to\jpsi\mathrm{K}^{*+}$}~and~\mbox{$\Bu\to\jpsi\mathrm{K}^{+}$}~decays.
  } \label{table:pt_pi11}
  \vspace*{3mm}
  \begin{tabular*}{0.65\textwidth}{@{\hspace{15mm}}c@{\extracolsep{\fill}}cc@{\hspace{15mm}}}
    \hline
    $\mathrm{p}^{\mathrm{T}}_{\piz}~~\left[\gevc\right]$
    & $\Peta^{\mathrm{corr}}_{\piz}~~\left[\%\right]$
    \\
    \hline 
    \medspace\medspace 0.5 \medspace\medspace -- 1.0 
    & $\medspace\medspace\medspace\medspace\medspace\medspace 93.9 \pm 7.8 \pm 2.5 \pm 6.1$
    \\
    \medspace\medspace\medspace\medspace 1.0 \medspace\medspace -- 1.25  
    & $\medspace\medspace\medspace\medspace\medspace\medspace 92.1 \pm 5.0 \pm 2.4 \pm 5.9$ 
    \\
    \medspace\medspace 1.25 -- 1.5 
    & $\medspace\medspace\medspace\medspace\medspace\medspace 94.9 \pm 4.3 \pm 2.3 \pm 6.1$ 
    \\
    \medspace\medspace 1.5 \medspace\medspace -- 2.0  
    & $\medspace\medspace\medspace\medspace\medspace\medspace 99.5 \pm 3.2 \pm 3.7 \pm 6.4$ 
    \\
    $\medspace\medspace\medspace\medspace\medspace\medspace\medspace\medspace\medspace\medspace > 2.0 $ 
    & $\medspace\medspace\medspace\medspace 117.2 \pm 3.7 \pm 2.6 \pm 7.6$
    \\ 
    \hline 
  \end{tabular*}
\end{table}

\begin{table}[h!tb]
  \centering
  \caption{
    Factors~$\Peta^{\mathrm{corr}}_{\piz}$~for 2012 data
    in bins of neutral pion transverse momentum. The first uncertainty is the statistical uncertainty,
    the second is systematic, and the third is related to an uncertainty
    of the known branching fractions of \mbox{$\Bu\to\jpsi\mathrm{K}^{*+}$}~and~\mbox{$\Bu\to\jpsi\mathrm{K}^{+}$}~decays.
  } \label{table:pt_pi12}
  \vspace*{3mm}
  \begin{tabular*}{0.65\textwidth}{@{\hspace{15mm}}c@{\extracolsep{\fill}}cc@{\hspace{15mm}}}
    \hline
    $\mathrm{p}^{\mathrm{T}}_{\piz}~~\left[\gevc\right]$
    & $\Peta^{\mathrm{corr}}_{\piz}~~\left[\%\right]$
    \\
    \hline 
    \medspace\medspace 0.5 \medspace\medspace -- 1.0 
    & $\medspace\medspace\medspace\medspace\medspace\medspace 89.7 \pm 5.9 \pm 1.6 \pm 5.8$
    \\
    \medspace\medspace\medspace\medspace 1.0 \medspace\medspace -- 1.25  
    & $\medspace\medspace\medspace\medspace\medspace\medspace 90.6 \pm 3.4 \pm 1.5 \pm 5.9$ 
    \\
    \medspace\medspace 1.25 -- 1.5 
    & $\medspace\medspace\medspace\medspace\medspace\medspace 94.9 \pm 3.1 \pm 1.6 \pm 6.1$ 
    \\
    \medspace\medspace 1.5 \medspace\medspace -- 2.0  
    & $\medspace\medspace\medspace\medspace 104.8 \pm 2.3 \pm 1.5 \pm 6.8$ 
    \\
    $\medspace\medspace\medspace\medspace\medspace\medspace\medspace\medspace\medspace\medspace> 2.0 $ 
    & $\medspace\medspace\medspace\medspace 116.4 \pm 2.6 \pm 1.6 \pm 7.5$
    \\ 
    \hline 
  \end{tabular*}
\end{table}

\section{Summary}
\label{seq:summary}

Using data corresponding to 3 fb$^{-1}$, collected in 2011 and 2012 with the LHCb detector, the correction factors for the \piz~and~\g~reconstruction efficiency are obtained. Their values are determined through measurement of the relative yields of \mbox{$\Bu\to\jpsi{}\mathrm{K}^{*+}\left(\to\mathrm{K}^+\piz\right)$} versus \mbox{$\Bu\to\jpsi\mathrm{K}^{+}$}~events. The correction factor for the photon~reconstruction efficiency are interpreted as $\Peta^{\mathrm{corr}}_{\g} = \sqrt{\Peta^{\mathrm{corr}}_{\piz}}$. The dependency of the corrections factors on the photon transverse energy is studied. 

\section{Acknowledgments}
\label{sec:acknowledgments}

I would like to thank Ivan Belyaev, Victor Egorychev and Daria Savrina for the invaluable help on every step of the work and for careful reading and the useful comments on this manuscript.

\bibliographystyle{LHCb}
\bibliography{main}

\end{document}